\documentclass[a4paper]{article}

\usepackage[T1]{fontenc}       
\usepackage[utf8]{inputenc}
\usepackage{lmodern}
\usepackage{bm} 
\usepackage{amsmath, amsfonts, amssymb}
\usepackage{graphicx}
\usepackage{hyperref}
\usepackage{authblk}
\usepackage{upgreek}

\newcommand{\cm}{cm$^{-1}$}

\newcommand{\Bps}{$B_{\mathrm{ps}}$}

\title{Large intravalley scattering due to pseudo-magnetic fields in crumpled graphene}

\author[1]{P\'{e}ter Kun}
\author[2]{Gerg\H{o} Kukucska}
\author[1]{Gergely Dobrik}
\author[2]{J\'{a}nos Koltai}
\author[2]{Jen\H{o} K\"{u}rti}
\author[1]{L\'{a}szl\'{o} P\'{e}ter Bir\'{o}}
\author[1]{Levente Tapaszt\'{o}}
\author[1, $\dagger$]{P\'{e}ter Nemes-Incze}

\affil[1]{Centre for Energy Research, Institute of Technical Physics and Materials Science, Budapest, Hungary}
\affil[2]{Department of Biological Physics, E\"{o}tv\"{o}s Lor\'{a}nd University (ELTE), Budapest, Hungary}
\affil[$\dagger$]{email: nemes.incze.peter@energia.mta.hu}

\date{}

\begin{document}

\maketitle

\section*{Abstract}

	The pseudo-magnetic field generated by mechanical strain in graphene can have dramatic consequences on the behavior of electrons and holes.
	Here we show that pseudo-magnetic field fluctuations present in crumpled graphene can induce significant intravalley scattering of charge carriers.
	We detect this by measuring the confocal Raman spectra of crumpled areas, where we observe an increase of the D'/D peak intensity ratio by up to a factor of 300.
	We reproduce our observations by numerical calculation of the double resonant Raman spectra and interpret the results as experimental evidence of the phase shift suffered by Dirac charge carriers in the presence of a pseudo-magnetic field.
	This lifts the restriction on complete intravalley backscattering of Dirac fermions.

\section*{Introduction}

	With the discovery of the half integer quantum Hall effect in graphene \cite{Novoselov2005,Zhang2005b} and of topological materials, \cite{Konig2007,Hsieh2008} Berry phase effects have taken center stage in condensed matter research.
	In graphene and other crystals with a honeycomb structure, charge carriers have a sublattice and valley degree of freedom, described in a continuum Dirac model by a pseudospin \cite{Beenakker2008,Xu2014}.
	As a consequence of this spin-like property, electrons or holes belonging to the two inequivalent valleys (K and K', see Fig. \ref{fig:psspin}) acquire a Berry phase of $\pi$ and $-\pi$ respectively during a cyclotron orbit.
	The importance of pseudospin and the Berry phase are most striking when perturbations are smooth on the atomic scale, ie. sublattice symmetry still holds.
	In this case, scattering between the two valleys is suppressed and the pseudospin is conserved \cite{Beenakker2008}, leading to some important effects that are the hallmark of graphene, such as weak antilocalization \cite{Tikhonenko2009}, the half integer quantum Hall effect \cite{Novoselov2005,Zhang2005b} and Klein tunneling \cite{Beenakker2008,Rickhaus2015}.
	Importantly, complete backscattering from a state $\left|-\bm{q}\right\rangle$ to $\left|\bm{q}\right\rangle$ (see Fig. \ref{fig:psspin}) is forbidden, due to pseudospin conservation \cite{Beenakker2008} ($\bm{q}$ is the crystal momentum measured from the K point in the Brillouin zone).
	This was first shown and explained for metallic carbon nanotubes \cite{Ando1998,McEuen1999} and is crucial for the exceptional mobility of graphene \cite{Banszerus2015a,Beenakker2008}.

	Here we show that scattering on strain fluctuations in graphene can lift the restriction on complete backscattering.
	This is demonstrated by confocal Raman spectroscopy measurements of crumpled graphene, where we measure a giant increase in the D' peak intensity.
	This Raman peak originates from a resonant Raman process which involves intravalley backscattering of charge carriers.
	The intravalley to intervalley scattering peak intensity ratio is found to be as high as $I_{\mathrm{D'}}/I_{\mathrm{D}} \approx 30$, in contrast to the usual value of $\approx$0.1 \cite{Eckmann2012}.
	Since the strain induced pseudo-magnetic field (\Bps) couples to the pseudospin \cite{Georgi2016a,Sasaki2008}, the enhancement of the D' peak at 1620 cm$^{-1}$ is due to the extra phase acquired by charge carriers undergoing Raman scattering on strain fluctuations.
	Thus, in contrast to a scalar potential, backscattering of Dirac particles is no longer forbidden.
	We reproduce our measurement results, using numerical calculation of the double resonant Raman processes.

\section*{Results and Discussion}

	Mechanical deformations in 2D materials with a honeycomb atomic structure naturally give rise to a two component pseudogauge field, which is directly proportional to the strain tensor components \cite{Kane1997,Suzuura2002,Kim2008c,Vozmediano2010,Cazalilla2014}.
	These strain induced fields have a scalar ($V(\bm{r})$) and a vector ($\bm{A}(\bm{r})$) component \cite{Vozmediano2010}, being analogous to an electrostatic potential and a magnetic vector potential.
	The latter having opposite sign in the two valleys and giving rise to a pseudo-magnetic field (\Bps) \cite{Vozmediano2010}.
	For graphene supported on hexagonal BN, \Bps\ is especially strong near bubbles (hundreds of Tesla) and has a major influence on transport properties \cite{Leconte2017}.
	Furthermore, in the highest mobility heterostructure devices it is very likely that random strain fluctuations are the main factors limiting mobility through intravalley scattering \cite{Couto2014}.
	This type of scattering is characterized by small changes in the charge carrier momentum and dominates if the scattering potentials are smooth on the atomic scale, such as charged impurities \cite{Casiraghi2007,Gibertini2012,Samaddar2015}, or strain fluctuations \cite{Couto2014}.
	Such scattering processes are mostly explored in charge transport experiments through weak (anti)localization measurements \cite{McEuen1999,Morozov2006,Tikhonenko2009}.
	In supported graphene, \Bps\ appears due to random strain fluctuations, stemming from non perfect stacking and interaction with the substrate.
	For SiO$_2$ supported graphene, \Bps\ due to corrugation has values of the order of 1T.
	The extra phase aquired by the wave function of the scattered carrier, much like a real magnetic field, suppresses the weak anti-localization effect \cite{Morozov2006}.

	Confocal Raman spectroscopy is another powerful tool to investigate strain fluctuations in graphene.
	Mechanical deformation induced softening or hardening of the phonon mode energy is detectable through the shift, splitting and broadening of the G and 2D peaks \cite{Couto2014,Neumann2015,Mueller2017}.
	However, until now direct detection of the scattering on strain fluctuations has been lacking.
	Both small (intravalley) and large momentum (intervalley) scattering is measurable separately via the double resonant D' and D peaks at $\approx$1620 cm$^{-1}$ and $\approx$1350 cm$^{-1}$.
	Lattice defects produce both intra- as well as intervalley scattering, giving contribution to both D' and D peaks, smooth defects which are less efficient at producing large momentum change, mostly contribute to the D' peak.
	Numerical calculations by Venezuela et al. \cite{Venezuela2011}, show that for closely packed alkali metals on graphene, acting as smooth Coulomb scattering centers, the D' peak should be more intense than the D peak.
	However, the peak itself should be undetectably small.
	Furthermore, for lattice defects, able to induce both types of scattering, the measured ratio of intensities is $I_{\mathrm{D'}}/I_{\mathrm{D}} \approx 0.1$ \cite{Eckmann2012}.
	This is much smaller than the expected value of $\sim$0.5 from analytical theory of the double resonant processes, suggesting that pseudospin effects could play an important role, particularly in intravalley processes \cite{Rodriguez-Nieva2014}.

\begin{figure}[h!]
	\includegraphics[width = 0.6 \textwidth]{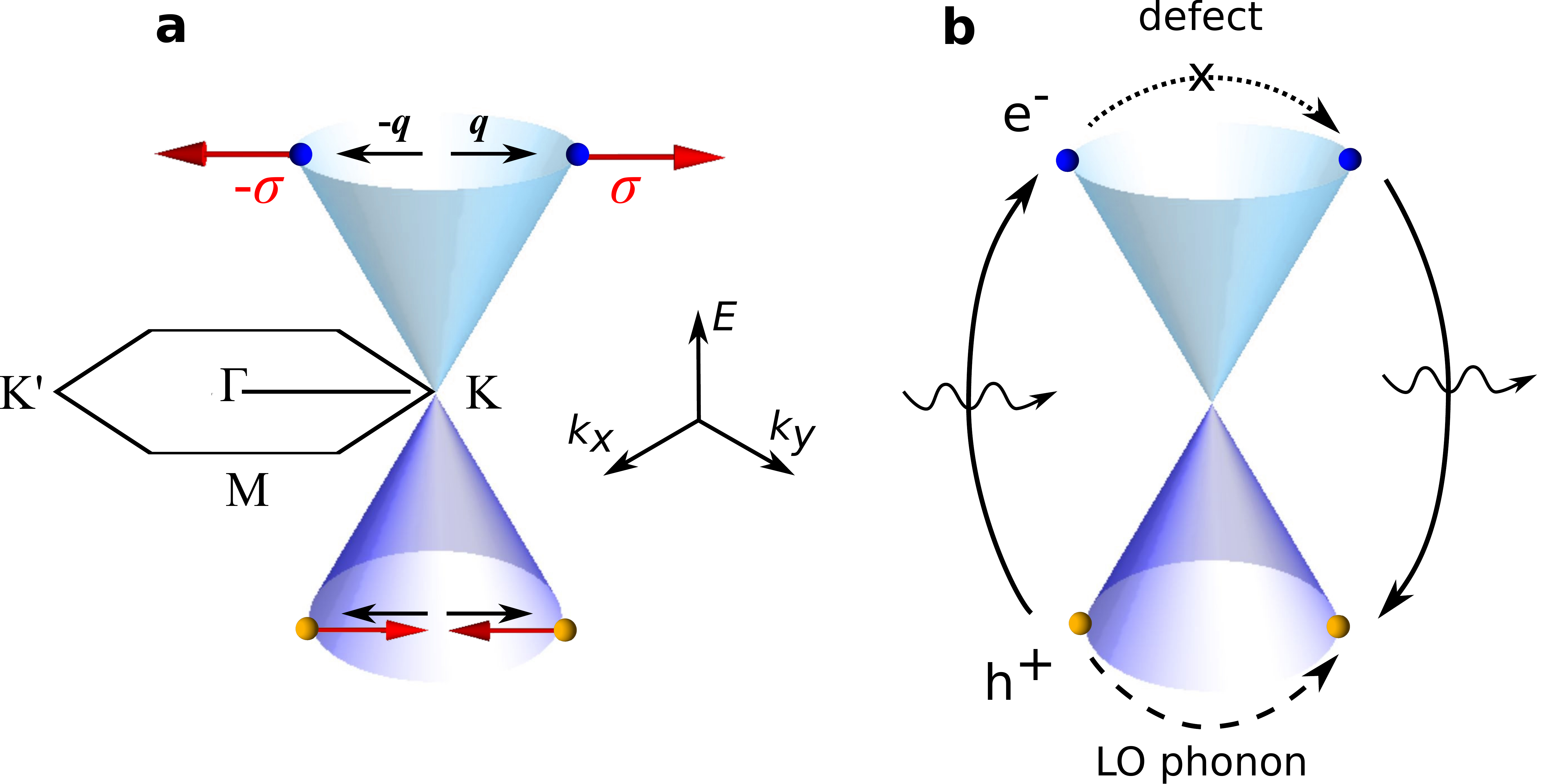}
	\caption{
		\textbf{Pseudospin and intravalley scattering in graphene}
		\textbf{(a)} Pseudospin texture around a K point in the first Brillouin zone of graphene.
		If the perturbation doesn't distinguish between the two sublattices, it is unable to flip the pseudospin ($\sigma$).
		Therefore, complete backscattering involving small changes in momentum from $\left|-\bm{q}\right\rangle$ to $\left|\bm{q}\right\rangle$ is forbidden by pseudospin conservation.
		\textbf{(b)} Intravalley process responsible for the D' Raman band of graphene.
	}
	\label{fig:psspin}
\end{figure}

	The double resonant D' process is sketched in Fig. \ref{fig:psspin}b.
	After the creation of the electron/hole pair the largest contribution to the D' intensity is given by processes involving scattering by both an electron and a hole \cite{Maultzsch2004,Venezuela2011}.
	Backscattering of the electron and/or hole involve an elastic defect scattering with small momentum change and a scattering involving an LO phonon along the $\overline{\mathrm{\Gamma M}}$ direction \cite{Malard2009,Venezuela2011}.
	It has been shown previously that the process involves a small portion of phonon phase space, as well as relatively small regions of the Dirac cone \cite{Maultzsch2004,Venezuela2011}.
	Since the biggest contribution to the intensity involves backscattering within a single valley along $\overline{\mathrm{\Gamma M}}$, this process necessarily involves a flip in pseudospin.
	Indeed it is suspected by Rodriguez-Nieva et al \cite{Rodriguez-Nieva2014} that the pseudospin related phases of the excited electron and hole play a dominant role in suppressing the backscattering necessary for the D' peak.
	Since Coulomb scatterers do not change the phase of the charge carrier wave functions, the resulting suppression of backscattering straightforwardly explains the immeasurably small calculated intensity of the D' peak \cite{Venezuela2011}.
	The situation changes drastically if the scattering potential has a vector component, ie. there is a sizable pseudo-magnetic field involved in the defect scattering.
	If the charge carriers stay within the same valley, this changes the phase of the electron or hole, similarly to a real magnetic field, enabling backscattering.

\begin{figure}
	\includegraphics[width = \textwidth]{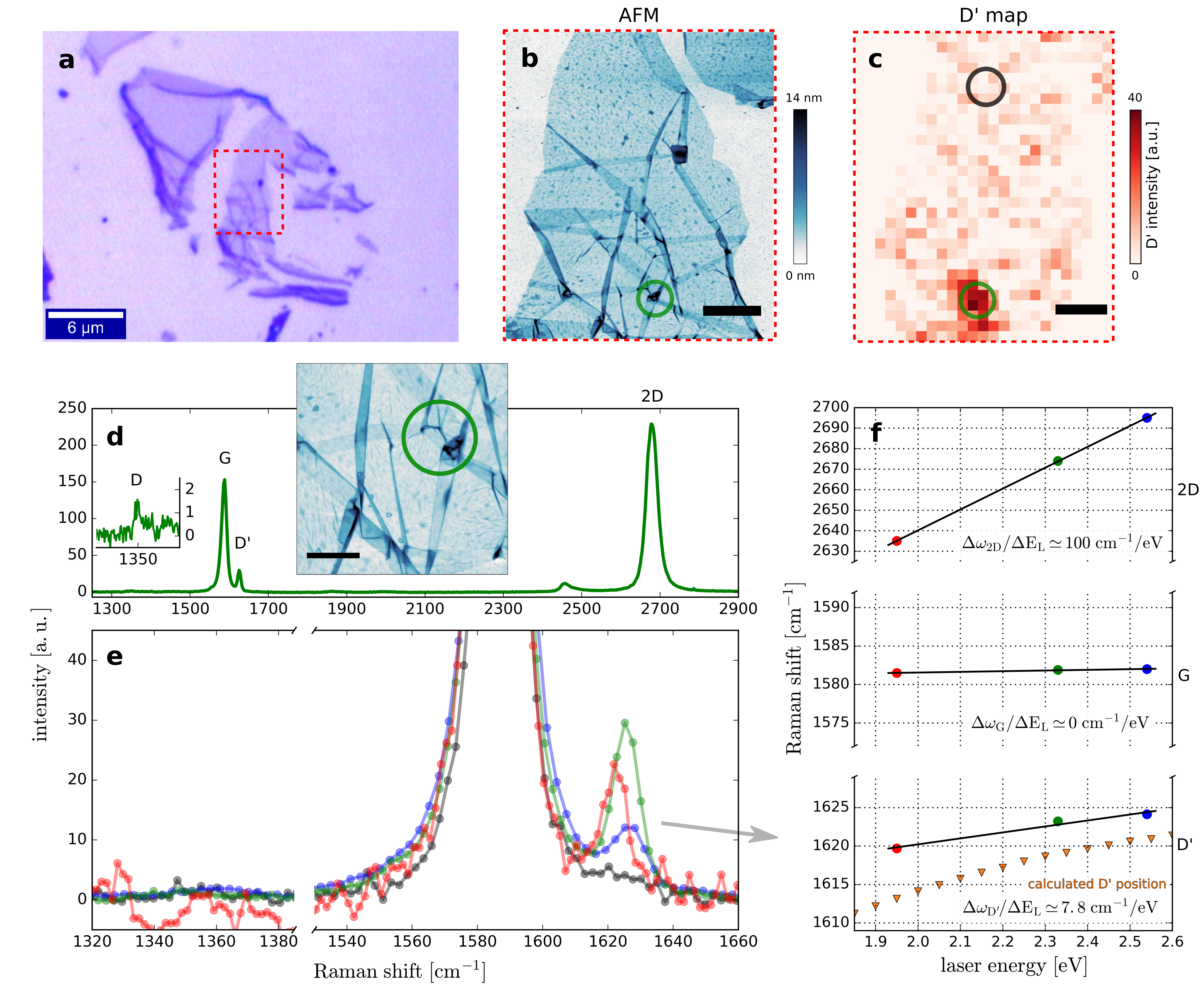}
	\caption{
		\textbf{Intravalley scattering in crumpled graphene.}
		\textbf{(a)} Optical microscope image of an exfoliated graphene layer on SiO$_2$.
		\textbf{(b)} AFM image of crumpled area. Position within flake shown by red rectangle in (a).
		\textbf{(c)} Raman spectroscopy map of the D' peak intensity in a crumpled graphene area in the position shown by the red rectangle.
		\textbf{(d)} Raman spectrum showing a large D' peak measured at the crumpled area shown in (b) and (c) by green circle, Insets: larger magnification AFM image of the crumpled area
		and zoom of the D peak region of the spectrum.
		\textbf{(e)} D' peak of the area shown in (d), measured with three different excitation wavelengths. Colors correspond to the respective laser color (red: 633 nm, green: 532 nm, blue: 488 nm). Black spectrum is measured using 532 nm excitation on the uncrumpled part of the flake, shown by the black circle in (c).
		\textbf{(f)} Change in the peak position ($\Delta \omega$) of the 2D, G and D' peaks, as a function excitation energy ($\mathrm{E_{L}}$), for the area marked by the green circle in (b, c, d).
		Measured dispersions ($\mathrm{\Delta \omega_{D^{\prime}} / \Delta E_{L}}$) of the experimental peaks are shown.
		Calculated D' peak position as a function of laser energy is shown by orange triangles.
	}
	\label{fig:experiment}
\end{figure}

	To detect Raman scattering from strain fluctuations with a sizable \Bps, we have measured confocal Raman maps on crumpled graphene flakes exfoliated onto a SiO$_2$ surface.
	We have used to our advantage that during the exfoliation some flakes tend to crumple (see Fig. \ref{fig:experiment}a). 
	Crumpling enhances \Bps\ caused by the strain fluctuations by at least a factor of 100 compared to the $\sim$1 Tesla \cite{Morozov2006} resulting from surface roughness of SiO$_2$.
	Additionally, we found the same results on samples that were crumpled using mild annealing and crumpling by a tungsten tip (see Methods).
	We have measured a total of ten crumpled graphene samples, among which six show an anomalously high and dispersive D' peak (see Supplementary information).
	Here, we illustrate the effect through the example of a characteristic sample (Fig. \ref{fig:experiment}), while data for other samples can be found in the Supplementary information.

\begin{figure}[h!]
	\includegraphics[width = 0.5 \textwidth]{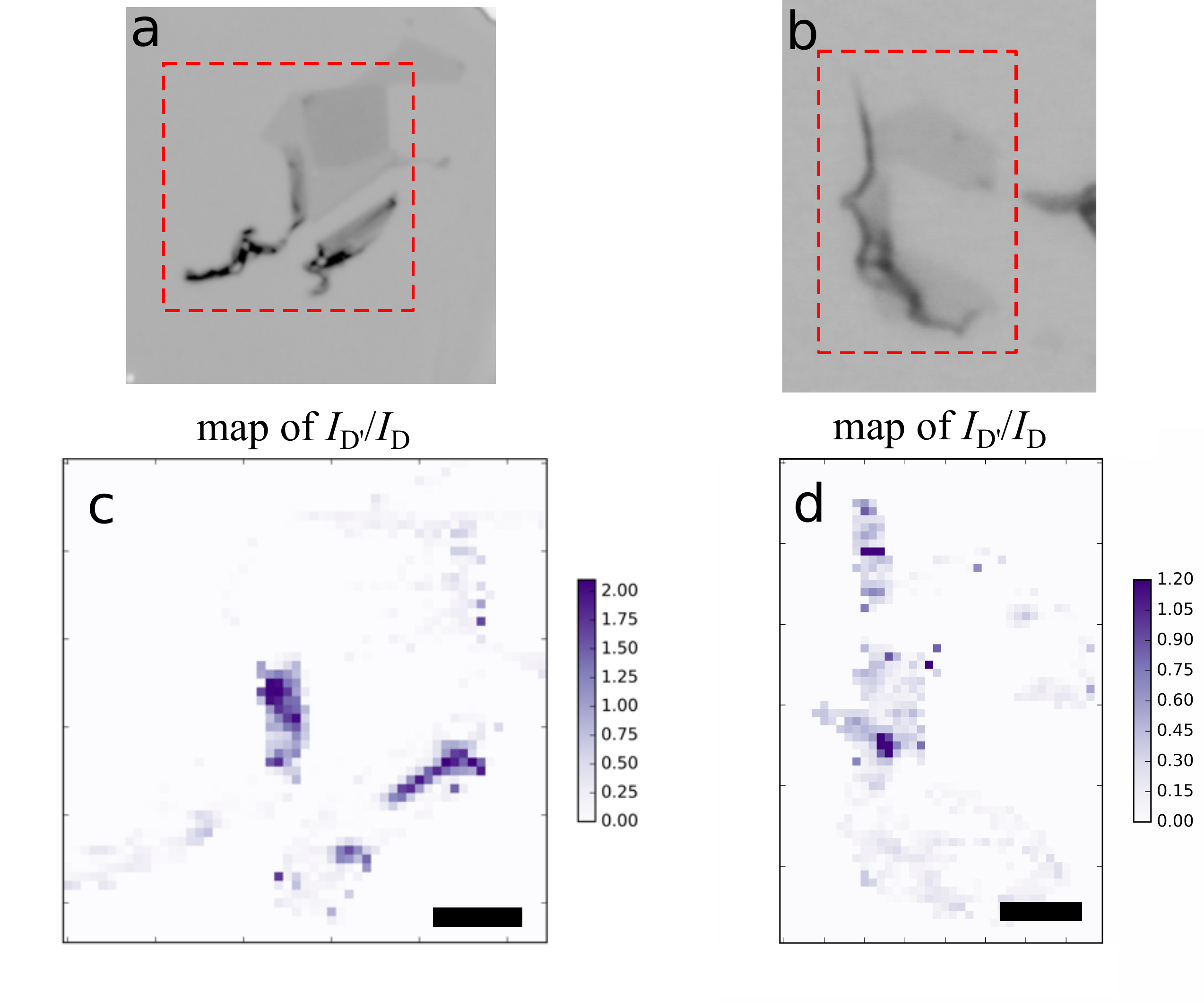}
	\caption{
		\textbf{$\bm{I_{\mathrm{D'}}/I_{\mathrm{D}}}$ intensity maps of crumpled graphene.}
		\textbf{(a, b)} Optical microscope images of two different crumpled graphene layers on SiO$_2$.
		\textbf{(c)} Map of the $I_{\mathrm{D'}}/I_{\mathrm{D}}$ ratio of the sample shown in (a).
		\textbf{(d)} Map of the $I_{\mathrm{D'}}/I_{\mathrm{D}}$ ratio of the sample shown in (b).
		Maps are measured using 532 nm excitation.
		Additional data regarding the samples and further samples can be found in the supplementary information.
	}
	\label{fig:experiment2}
\end{figure}

	AFM measurements show that the graphene layer consists of a network of folds on it's surface.
	Measuring the Raman spectra with a spatial resolution of $\sim$500 nm and excitation wavelength of 532 nm, we map the intensity of the D' peak (Fig. \ref{fig:experiment}c) within the crumpled area marked in Fig. \ref{fig:experiment}a.
	The most striking feature of the map is that the crumpled regions show increased D' intensity, the region with the strongest enhancement marked by a green circle.
	To shed more light on the Raman scattering, we plot the complete Raman spectrum of this region in Fig. \ref{fig:experiment}d.
	The most surprising feature of the spectrum is the extremely high D' peak, having 20\%\ the intensity of the G peak.
	This is unprecedented because the D' peak is only observed when the sample contains strong lattice defects, such as vacancies, \emph{sp$^3$} type defects of grain boundaries \cite{Pimenta2007,Venezuela2011,Eckmann2012}, and it is always accompanied by the D peak, with its intensity being mostly $\sim$10\%\ of the D intensity ($I_{\mathrm{D'}}/I_{\mathrm{D}} \approx 0.1$) \cite{Eckmann2012,Eckmann2013}.
	This result has to be considered in the light that in the area where the spectrum was measured there are no graphene edges.
	As expected for defect free graphene, the D peak intensity is just barely larger than the background (see inset in Fig \ref{fig:experiment}d).
	Furthermore, the graphene flake shows the pristine Raman spectrum of graphene in the uncrumpled areas.
	Ratios $I_{\mathrm{D'}}/I_{\mathrm{D}} \approx 10$ have been also found in samples where the laser spot contains both crumpled areas and edges (see Supplementary information).
	The $I_{\mathrm{D'}}/I_{\mathrm{D}}$ maps of two additional samples can be seen in Fig. \ref{fig:experiment2}.

	It is known that overlapping graphene layers can produce a non dispersive peak ranging from 1540 up to 1630 \cm\ \cite{Lu2013,Carozo2011,Carozo2013}.
	The non dispersive nature of this peak is due to the fact that the LO phonon from around the $\Gamma$ point taking part in the process is selected by the mismatch angle of the two graphene layers and not by the laser energy \cite{Carozo2011}.
	Therefore, if the mismatch angle between the overlapping graphene layers is in the 4$^{\circ}$ to 6$^{\circ}$ range, this so called R' peak could be mistaken for the D' peak.
	To make sure that the peak around 1620 \cm\ is indeed the D' peak, we measured the change in the peak position with changing excitation wavelength.
	The data for the presented sample can be seen in Fig. \ref{fig:experiment}e, with the graph color corresponding to the excitation laser color.
	The measured dispersion is 7.8 $\mathrm{cm}^{-1}/\mathrm{eV}$ and is similar to the values measured on other samples (see Supplementary information).
	It is slightly lower than the value of $\sim$10 $\mathrm{cm}^{-1}/\mathrm{eV}$ expected for the D' peak \cite{Pimenta2007,Eckmann2013}, see also our calculation of the strain induced D' peak dispersion in the same plot (see Fig. \ref{fig:experiment}f).
	The reason for the lower dispersion value might be the fact that the D' peak is mixed with the non dispersive R' peak in certain areas.
	An example where this effect is strong is shown in Figure S5 of the Supplementary information.
	To reduce the possible interference from the R' peak in our measurements, we have prepared an additional set of samples.
	These graphene samples had their top and bottom covered in a 5 nm thick poly-vinyl alcohol (PVA) film.
	Crumpling was achieved by using a sharp tungsten needle and a micro-manipulator stage.
	We found similar D' peak intensities in the crumpled areas as in the case of other samples (see Supplementary figures S6 and S7).
	These experiments rule out any major effect from the R' peak, since the PVA layer on the graphene prevents any atomically clean interface to form between the overlapping graphene.

	Within the literature there have been some observations of a barely measurable D' peak on graphene supported on nanosized pillars \cite{Pacakova2017} and nanoparticles \cite{Vejpravova2015}, exhibiting wrinkling.
	However, it is unclear what the D to D' intensity ratio is in these experiments.
	In experiments of graphene wrinkling on a polymer, the D' peak is obscured by the presence of the polymer substrate \cite{Androulidakis2017}.

\begin{figure}[h]
	\includegraphics[width = \textwidth]{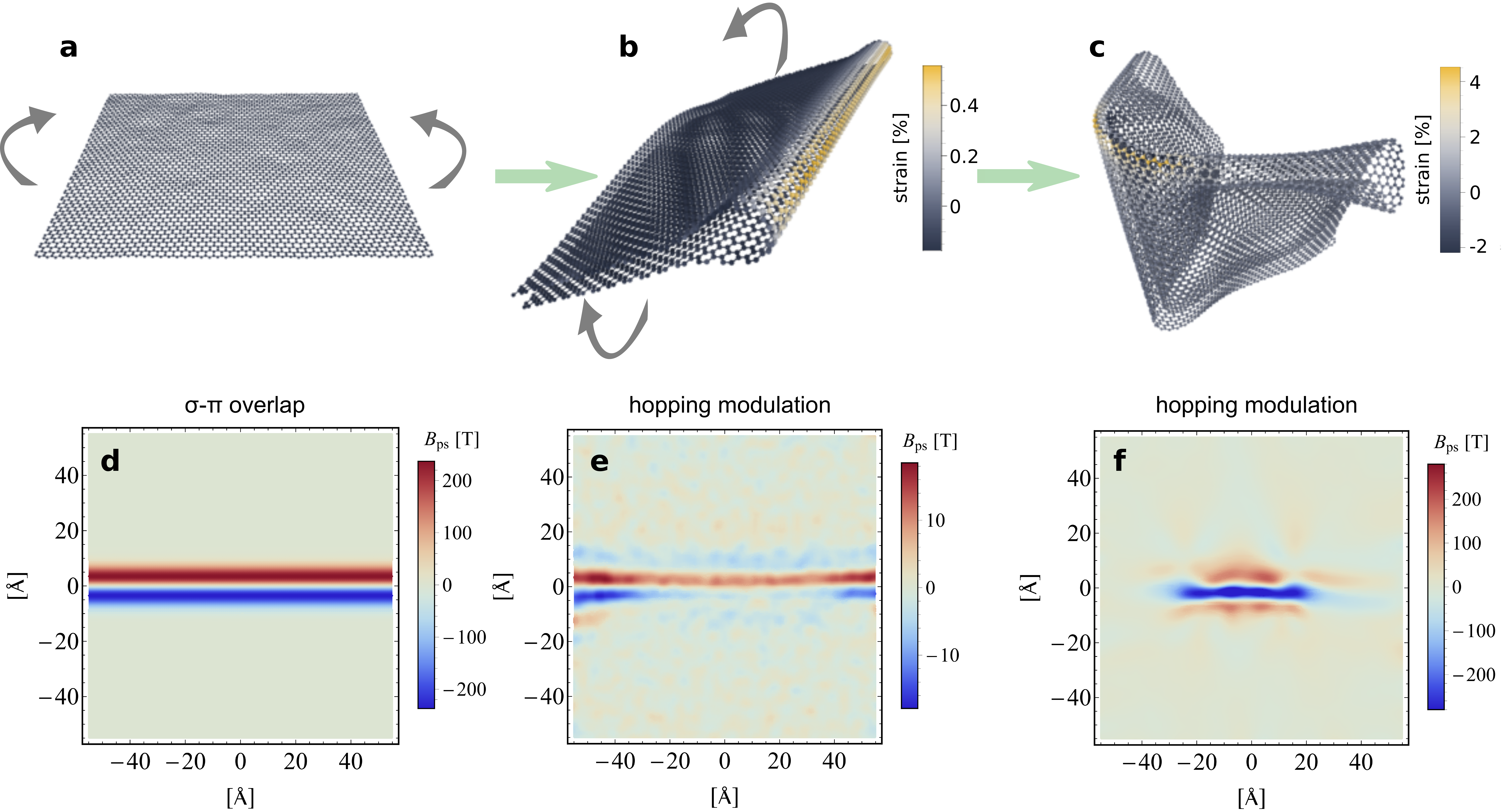}
	\caption{
		\textbf{Strain within graphene folds, deformations.}
		\textbf{(a)} 10 $\times$ 10 nm graphene piece used in the molecular dynamics calculations \cite{Plimpton1995}.
		\textbf{(b)} Single graphene fold along the zigzag direction.
		\textbf{(c)} Doubly folded graphene.
		Color scale on atoms shows the local strain, calculated by taking the average of the relative nearest neighbor distance changes. Positive values (yellow) correspond to tensile strain.
		\textbf{(d-f)} Pseudo-magnetic field magnitude in the structures shown in (b) and (c).
		\textbf{(d)} Pseudo-magnetic field generated by the $\sigma$-$\pi$ overlap \cite{Kim2008c} within the fold seen in (b).
		\textbf{(e)} \Bps\ magnitude due to the hopping modulation within the fold seen in (b).
		\textbf{(f)} \Bps\ magnitude due to hopping modulation within the double fold seen in (c)
	}
	\label{fig:strain}
\end{figure}

	Next, we reproduce our measured D' peak intensities by numerical calculations of the double resonant processes \cite{Kurti2002,Venezuela2011}.
	To explore the pseudo-magnetic field within crumpled graphene and to construct a theoretical model for the Raman scattering potential, we study the wrinkling of graphene, using the LAMMPS molecular dynamics code \cite{Plimpton1995} (for details see Supplementary information).
	We consider two model geometries that make up a crumpled graphene sheet \cite{Vliegenthart2006}.
	One involves a single fold (Fig. \ref{fig:strain}b), seen all over the crumpled sample (Fig. \ref{fig:experiment}b), the other involves a double fold (Fig. \ref{fig:strain}c), constructed by creating a second fold in a singly folded graphene.
	The double fold shows a complex strain pattern, containing both compression and tension and an increase in the strain by a factor of 10 compared to the single fold, as evidenced by the color scale on the atoms in Fig. \ref{fig:strain}b,c.
	
	Having determined the geometry of the crumpled graphene, it becomes possible to calculate the the pseudo-magnetic field \Bps, from the atomic positions.
	It can be understood, as a result of the local modification of inter-atomic hopping, either through changes in bond length or bond angles \cite{Vozmediano2010}.
	Additionally, \Bps\ can be interpreted as the consequence of Dirac quasiparticles existing in a curved space-time \cite{Vozmediano2010,DeJuan2007,Arias2015,Ochoa2016,Castro-Villarreal2017}.
	Here, we calculate \Bps\ from the modulation of the hopping parameter, as described by Guinea et al. \cite{Guinea2009}.
	This model reproduces the magnitude of \Bps\ measured via scanning tunneling microscopy \cite{Levy2010,Georgi2016a,Jiang2017c}.
	Computing \Bps, directly from the modulation of the atomic positions \cite{Barraza-Lopez2013} we find a maximum \Bps\ of 20 T in the single fold, and as expected from the difference in strain, the double fold shows a \Bps\ in the 200 T range (Fig. \ref{fig:strain}e,f).
	This \Bps\ originates from the modulation of the nearest neighbor hopping energy due to strain \cite{Kane1997,Suzuura2002,Guinea2009}.
	However, in cases where the curvature of the graphene sheet is large, there is another sizable component to the pseudo-magnetic field, due to the hybridization of the $\sigma$ and $\pi$ bonds \cite{Kim2008c}.
	The total vector potential is the sum of the hopping induced $\bm{A}(\bm{r})$ and the curvature induced $\bm{A}^{\sigma \pi}(\bm{r})$.
	The experimentally relevant part of the vector potential \cite{Georgi2016a}, the pseudo-magnetic field, is then formed by the rotor of the sum of these two contributions: $B_{\mathrm{ps}} = \nabla \times (\bm{A} + \bm{A}^{\sigma \pi})$.
	Using the formula for $\bm{A}^{\sigma \pi}$ from Rainis et al \cite{Rainis2011}, we calculate the \Bps\ for our fold.
	For a radius of curvature $R$ of a fold parallel to the zigzag direction, we have $A^{\sigma \pi}_x(\bm{r}) = 3 \varepsilon_{\pi \pi} a^2 / 8 R(\bm{r})^2$ and $A^{\sigma \pi}_y(\bm{r}) = 0$, where $\varepsilon_{\pi \pi} \approx 3\ \mathrm{eV}$ \cite{Rainis2011,Kim2008c} and $a$ = 1.42 \AA.
	In our calculations $R$ is around 3 \AA, corresponding to the 2-3 \AA\ measured by Annett et al. \cite{Annett2016}, while Rainis et al \cite{Rainis2011} calculate 7 \AA.
	As an example, in the case when $R$ = 4 \AA\ the maximum curvature induced \Bps\ is around 200 T, an order of magnitude larger than the hopping induced one (see Fig. \ref{fig:strain}d,e), in accordance with the finding of Rainis et al \cite{Rainis2011}.
	Both contributions to \Bps\ are also dependent on the orientation of the fold within the graphene lattice, with the maximum \Bps\ present in folds parallel to the zigzag direction and zero \Bps\ for armchair \cite{Rainis2011,Carrillo-Bastos2016}.
	For the double fold, the hopping induced \Bps\ is in itself in the 200 Tesla range, due to the larger strains (Fig. \ref{fig:strain}f).
	A large collection of double or multiple folds may be necessary to create the large \Bps\ values needed to observe the enhanced D' peak, such as in the region shown by a green circle in Fig. \ref{fig:experiment}b,c.
	Using the field theoretical approach in curved space \cite{DeJuan2007,Arias2015,Ochoa2016,Castro-Villarreal2017} to calculate \Bps, we find values of similar magnitude.
	However, in our experiment we lack the spatial resolution to be able to distinguish between the two possible interpretations of \Bps\ \cite{Liu2018b}.
	For further discussion see section S6 of the Supplementary information.
	
	We mention that here we assume \Bps\ is not homogeneous on the scale of the magnetic length, thus Landau quantization \cite{Levy2010} due to strain is not expected.
	As an example, the magnetic length at 200 T is $l_B \approx 1.7$ nm.
	This length scale is more than a factor of 3 larger than the size of areas with high magnetic field, see Fig. \ref{fig:strain}f, making the formation of Landau levels impossible.
	Even though Landau levels are not expected to form, it has been shown that folded graphene areas with rapidly changing $B_{\mathrm{ps}}$ can host bound states of Dirac fermions, as shown experimentally \cite{Wu2018a} and theoretically \cite{Wakker2011,Zhang2014b}.

\begin{figure}[h]
	\includegraphics[width = \textwidth]{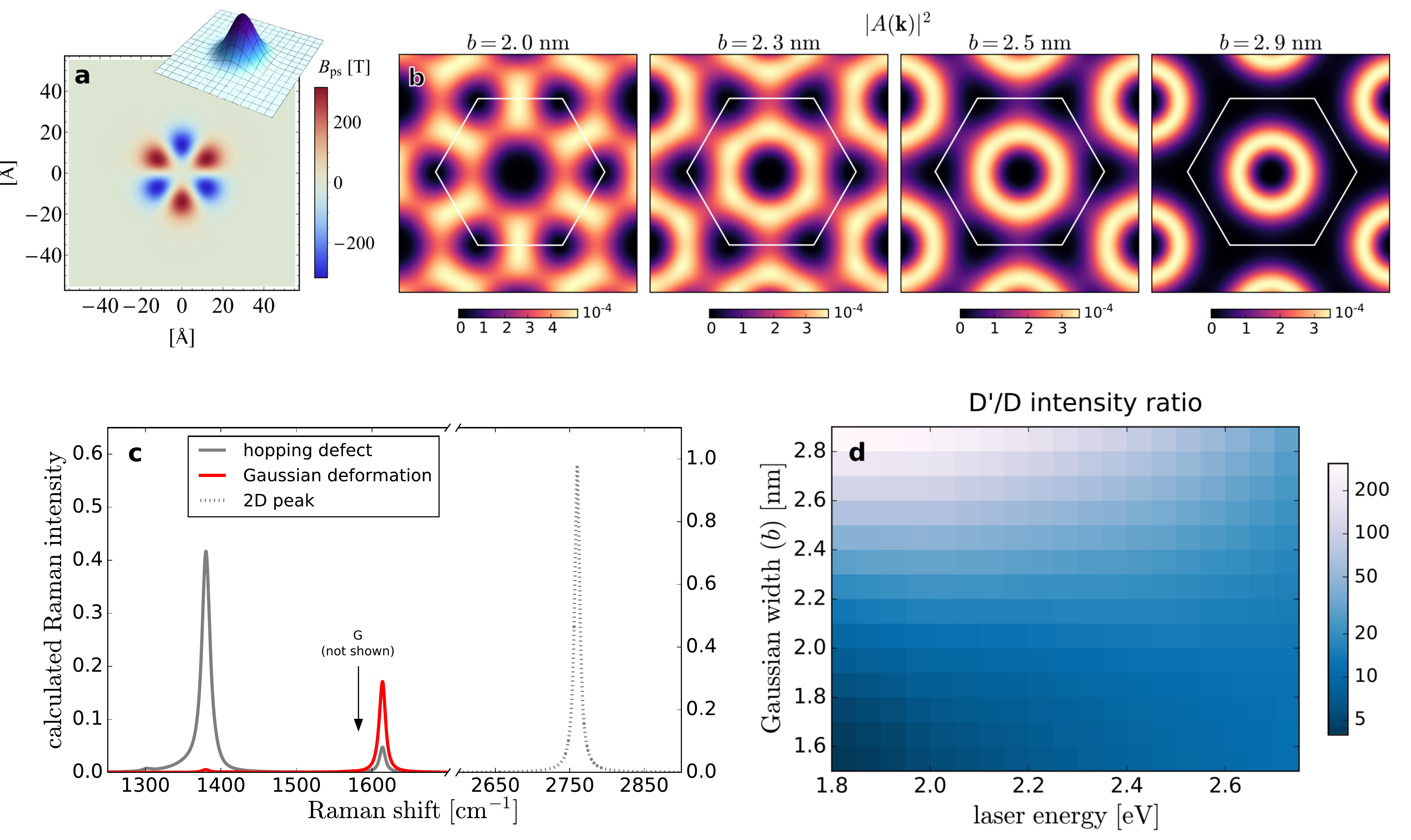}
	\caption{
		\textbf{Calculated resonant Raman spectra.}
		\textbf{(a)} Model used in Raman calculations. Gaussian of shape: $h \cdot \mathrm{exp}(-\frac{x^2 + y^2}{b^2})$, where $h$ is the height and $b$ is the halfwidth of the deformation (inset). Color plot shows the calculated pseudo-magnetic field pattern for a deformation having $b$ = 2 nm and $h$ = 0.5 nm.
		\textbf{(b)} Plot of $|\bm{A}(\bm{k})|^2$ in the first Brillouin zone (white hexagon).
		\textbf{(c)} Calculated D and D' peak intensity for a Gaussian deformation seen in (a).
		\textbf{(d)} Ratio of D' and D intensities ($I_{\mathrm{D'}}/I_{\mathrm{D}}$) as a function of excitation energy and Gaussian width $b$. The Gaussian amplitude is $h = b/4$, in order to keep the magnitude of the mechanical strain within the Gaussian constant, only changing the spatial extent of the deformation. Color scale is logarithmic.
	}
	\label{fig:calc}
\end{figure}

	Having quantified the \Bps\ magnitude in folded, crumpled graphene, we turn our attention to numerically calculating the double resonant Raman processes in the presence of strain.
	Calculations are performed similarly to Venezuela et al \cite{Venezuela2011} and K\"{u}rti et al \cite{Kurti2002}.
	For this we need to find the electron/hole - defect Hamiltonian: $H_{\mathrm{\mathrm{def}}}$.
	We model the \Bps\ seen in crumpled graphene by a simple model, using a Gaussian deformation of the form: $h \cdot \mathrm{exp}(-\frac{x^2 + y^2}{b^2})$, where $h$ is the height and $b$ is the width of the Gaussian.
	For a Gaussian, the analytical form of the vector potential $|\bm{A}| \propto h^2/b^2$ is well known \cite{Schneider2015}.
	By choosing the height $h$ and width $b$ of the bump to be 5 \AA\ and 20 \AA\ respectively, the \Bps\ within the bump area reaches a maximum of 200 T, switching sign on the nanometer scale (see Fig. \ref{fig:calc}a).
	Both the magnitude and the length scale is similar to the values found within the LAMMPS calculations.
	In atomic units, the Hamiltonian describing the defect is $H_{\mathrm{def}} = \bm{A}(\bm{r}) \cdot \nabla + V(\bm{r})$, where $\bm{A}$ is the vector potential and we also include $V$, the scalar potential generated by the mechanical deformation \cite{Vozmediano2010}.
	In order to calculate the scattering matrix element $\left\langle \bm{k} \mid H_{\mathrm{def}} \mid \bm{k}\pm\bm{q} \right\rangle$ between two states with wave vectors $\bm{k}$ and $\bm{k}\pm\bm{q}$ we need to calculate $\bm{A}(\pm \bm{q})$ and $V(\pm \bm{q})$ \cite{Venezuela2011}:
	\begin{equation}
		\begin{aligned}
		V(\pm \bm{q}) = \sum_{l} e^{\pm i \bm{q} \bm{r}_l} V(\bm{r}_l) \\
		\bm{A}(\pm \bm{q}) = \sum_{l} e^{\pm i \bm{q} \bm{r}_l} \bm{A}(\bm{r}_l)
		\end{aligned}
	\label{eq:potentials}
	\end{equation}
	Fig. \ref{fig:calc}b shows $|\bm{A}(\bm{k})|^2$ in the first Brillouin zone of graphene for varying sizes of the deformation.
	While varying the width, the height to width ratio ($h/b$) was kept constant at 0.25, in order to keep the maximum value of $|\bm{A}(\bm{k})|$ constant \cite{Schneider2015}.

	The calculated D, D' and 2D peaks for a Gaussian with $h = 5$ \AA\ and $b = 20$ \AA\ is shown in Fig. \ref{fig:calc}c.
	We reproduce $I_{\mathrm{D'}}/I_{\mathrm{D}} = 32$ in accordance with the experimental spectrum of Fig. \ref{fig:experiment}d.
	As a crosscheck to our calculations we also compute the D and D' intensity for a "hopping" defect, as implemented by Venezuela et al \cite{Venezuela2011}, modeling a lattice defect in graphene.
	For the hopping defect, the D' peak intensity is 11\%\ of the D intensity, as expected for lattice defects, such as vacancies \cite{Eckmann2012}.
	Comparing the relative intensity with respect to the 2D peak, it is clear that the D' peak is showing a measurable intensity, as opposed to the D' peak generated by Coulomb scatterers \cite{Venezuela2011}.
	The D' intensity is generated almost completely by the vector potential $\bm{A}$, with the scalar potential giving only a slight (less then $10^{-4}$) contribution (see Supplementary information).
	This can be easily explained if we consider the scattering matrix element between states $|\left\langle \bm{k} \mid H_{\mathrm{def}} \mid \bm{k}\pm\bm{q} \right\rangle|^2 = |H_{\mathrm{def}}({\bm{q}})|^2 \mathrm{cos}^2(\theta_{\bm{k}, \bm{k}\pm\bm{q}}/2)$, where $\theta_{\bm{k}, \bm{k}\pm\bm{q}}$ is the angle between the the initial and the scattered state \cite{Ando1998}.
	For backscattering ($\theta_{\bm{k}, \bm{k}\pm\bm{q}} = \pi$), if $H_{\mathrm{def}}$ is of purely scalar character, the matrix element is zero.
	However, if $H_{\mathrm{def}}$ has a vector potential component it can be nonzero.
	This is because $\bm{A}$ acts on the pseudospin of graphene, as evidenced by the pseudo-Zeeman effect \cite{Georgi2016a}.
	Thus, it is able to change the phase of the charge carriers.

	The vector potential $\bm{A}$ has a slight contribution to the D peak as well, as can be seen in Fig. \ref{fig:calc}c.
	This is due to the non zero scattering potential at large $\bm{k}$ values, around the K points.
	Due to the Fourier transformation in eq. \ref{eq:potentials}, the contribution at large $\bm{k}$ increases as we make the Gaussian narrower.
	By decreasing $b$, but keeping the aspect ratio the same, we can see a marked increase in the scattering potential $|\bm{A}(\bm{k})|^2$ at the K points, increasing the D peak intensity.
	This can be observed in Fig. \ref{fig:calc}d, where we plot the $I_{\mathrm{D'}}/I_{\mathrm{D}}$ ratio on a log scale, as a function of $b$ and excitation energy.
	Although it is clear that the parameter $b$ is a tuning knob for the $I_{\mathrm{D'}}/I_{\mathrm{D}}$ ratio, it is not possible to attribute a certain $b$ value to the measured ratio, because the crumpling patterns can have a more complex \Bps\ pattern than what is assumed in the Gaussian model.
	Additionally, the calculated model does not include lattice defects, which could be present in the experiment, such as edges.
	For the map seen in Fig. \ref{fig:experiment}b, the D peak intensity is too low to prepare a meaningful $I_{\mathrm{D'}}/I_{\mathrm{D}}$ map.
	For other samples, the map of this ratio can be found in the Supplementary information.

	Our experiments show a first example of resonant Raman scattering on potentials that are smooth on the atomic scale.
	In the lack of intervalley scattering, the pseudo-magnetic field created by the strain induced vector potential shifts the phase of the electron or hole wave function, in addition to the Berry phase.
	This lifts the restriction on backscattering, enabling an enhancement in the intravalley backscattering rate by orders of magnitude, evidenced by the enhancement of the D' peak in the Raman spectrum.
	The drastic increase in intravalley scattering, as a consequence of strain fluctuations, may be an important factor in lowering the mobility of certain CVD grown graphene samples, especially if wrinkles and folds are introduced during the transfer process.
	Furthermore, it is well known that defects such as grain boundaries, vacancies, etc. can have a specific $I_{\mathrm{D'}}/I_{\mathrm{D}}$ fingerprint \cite{Eckmann2012}.
	Such defects also distort the graphene lattice around them \cite{Rasool2014} and this local strain field can be specific to the type of defect.
	Our results show that if we want to understand the origin of this Raman fingerprint, scattering on the defect induced local strain fields has to be taken into account.
	If experimentally, a more controlled folding of graphene can be achieved, the \Bps\ present in these folds could be used to steer and guide \cite{Wu2018a} valley polarized \cite{Wu2011,Jiang2013} Dirac fermions.

\subsection*{Methods}

	Graphene samples were prepared by micromechanical exfoliation from natural graphite, purchased from NGS Trading \& Consulting GmbH.
	Raman measurements were carried out using a Witec 300rsa+ confocal Raman system, using 488 nm, 532 nm and 633 nm excitation lasers.
	Laser power was kept at 0.5 mW for all lasers.
	
	Crumpling of the graphene layers has been induced either in the exfoliation stage, as with the sample in the main text and the sample shown in Fig. S4 of the Supplementary information.
	However, this method has a low yield.
	It is known that annealing can induce folding of graphene. \cite{Annett2016}.
	Therefore, to increase the yield of the crumpled graphene within our samples, we have used annealing to induce wrinkling and crumpling.
	Sample 1 to 4 have been prepared by placing them onto a hotplate, preheated to 160$^{\circ}$C for 15 minutes, after which the samples are removed and placed onto a metal surface to allow quick cooling.
	Additionally, the crumpling can be also induced by the tip of a tungsten needle.

\subsection*{Data availability}
	The datasets generated during and/or analysed during the current study are available from the corresponding author on reasonable request.

\subsection*{Acknowledgement}

	We acknowledge valuable discussions with Nancy Sandler and Daiara Faria.
	Support from the Hungarian National Research, Development and  Innovation Office (NKFIH, Grants No. K-115608, K-108676) is  acknowledged.
	The work was conducted within the Graphene Flagship, H2020 GrapheneCore2 project no. 785219.
	This research was supported by the National Research Development and Innovation Office of Hungary within the Quantum Technology National Excellence Program (Project No. 2017-1.2.1-NKP-2017-00001).
	JKo and DG acknowledges the Bolyai program of the Hungarian Academy of Sciences.
	L.T. acknowledges financial support from OTKA grant K108753 and NanoFab2D ERC Starting grant.
	PNI acknowledges support form the Hungarian Academy of Sciences, within the framework of the MTA EK Lend\"{u}let Topology in Nanomaterials Research Group, grant no: LP2017-9/2017.

\subsection*{Additional information}
	Competing interests: the authors declare no competing financial or non-financial interests.

\subsection*{Author contributions}
	PK and GK contributed equally.
	PK carried out the experiments together with PN-I, the Raman calculations were carried out by GK under supervision from JK.
	DG prepared crumpled graphene samples.
	PN-I carried out the molecular dynamics calculations.
	PN-I conceived and coordinated the project, together with LT and LPB.
	PN-I wrote the manuscript, with contributions from all authors.


\end{document}